\documentclass{elsart}

\usepackage{graphicx}
\usepackage{natbib}  
\usepackage{amsmath}

\journal{Bulletin of Mathematical Biology}

\def\erf{\mathop{\rm erf}\nolimits}

\begin{document}

\begin{frontmatter}

\title{Diffusion and Home Range Parameters from  Rodent Population Measurements in Panama}

\author[consortium]{L. Giuggioli},
\ead{giuggiol@unm.edu}
\author[consortium,cab]{G. Abramson},
\ead{abramson@cab.cnea.gov.ar}
\author[consortium]{V. M. Kenkre}
\ead{kenkre@unm.edu}

\address[consortium]{Consortium of the Americas for Interdisciplinary Science,
University of New Mexico, Albuquerque, New Mexico 87131, USA.}

\address[cab]{Centro At\'{o}mico Bariloche, CONICET and Instituto Balseiro, 8400  San
Carlos de Bariloche, R\'{\i}o Negro, Argentina.}

\author{G. Suz\'{a}n},
\ead{gsuzan@unm.edu}
\author{E. Marc\'{e}},
\ead{emarce@unm.edu}
\author{T. L. Yates}
\ead{tyates@unm.edu}
\address{Department of Biology, University of New Mexico, Albuquerque, New
Mexico 87131, USA.}

\date{\today}

\begin{abstract}
Simple random walk considerations are used to interpret rodent  population data
collected in Hantavirus-related investigations in  Panama regarding the
short-tailed cane mouse, \emph{Zygodontomys brevicauda}. The diffusion constant
of mice is evaluated to be of the order of (and larger than) 200 meters squared
per day. The investigation also shows that the  rodent mean square displacement
saturates in time, indicating the  existence of a spatial scale which could, in
principle, be the home  range of the rodents. This home range is concluded to
be of the order of 70 meters. Theoretical analysis is provided for interpreting
animal movement data in terms of  an interplay of the home ranges, the
diffusion constant, and the size of the grid used to monitor the movement. The
study gives impetus to a substantial modification of existing theory of the
spread of the  Hantavirus epidemic which has been based on simple diffusive
motion of  the rodents, and additionally emphasizes the importance for
developing  more accurate techniques for the measurement of rodent movement.

\end{abstract}


\end{frontmatter}

\section{Introduction}

The Hantavirus epidemic is of great concern to human health in many regions of
the world \citep{yatesetal,mills99,parmenter99}. The discovery of Hantavirus in
the New World took place after an outbreak of a severe disease in the region of
the Four Corners in the North American Southwest, in 1993. The agent,
Hantavirus Sin Nombre, is carried mainly by the extremely common deer mouse,
\emph{Peromyscus maniculatus}~\citep{nichol93,childs94}. Since the discovery,
an enormous effort has been devoted to understanding the ecology and
epidemiology of the virus-mouse association, with the ultimate goal being
prediction  of human risk~\citep{mills99,cparmenter98}. Numerous species of the
virus are known in the Americas, each one of them almost exclusively associated
with a single rodent reservoir~\citep{schmaljohn97}. Human disease caused by
these pathogens can range from mild to very severe, with a mortality rate in
some cases approaching 50\%.

A theory for the spread of the Hantavirus was constructed a few years  ago by
two of the present authors~\citep{ak} and shown to lead naturally to
spatiotemporal patterns such as the observed refugia~\citep{yatesetal} and the
sporadic disappearance and appearance of the
epidemic~\citep{mills99,parmenter99}. That theory will be referred to in the
rest of the paper as AK. Seasonal, as well as extraordinary variations in
demographic and environmental conditions are included in the AK model through
spatio-temporal dependence of several parameters, such as the carrying
capacity. The emergence of traveling waves of infection~\citep{akpy}, the
investigation of fluctuations~\citep{aabk}, external changes in environmental
effects~\citep{bkk}, and other features~\citep{kpasi,apasi,kenkre04,buceta04}
have also been studied. There are additional factors, not yet analyzed
theoretically, which may be of importance in the dynamics of the virus. For
example, human activity such as changes in agricultural practice may alter
habitats and drive the rodent population into new habitats not previously
occupied by them. While many theoretical issues regarding the AK development
have been, and are being, explored quite intensely, the major problem of
obtaining the values of the parameters inherent in the theory has remained
neglected as a result of paucity of available data. Needless to say, the
solution of this problem is crucial to the quantitative description of the
spread of the epidemic. The purpose of the present  paper is such extraction of
the essential parameters necessary in the description of the spread of the
Hantavirus.

\begin{table}[b]
\caption{Summary of the mark-recapture data of \emph{Z. brevicauda}.  A total
of 846 captures, corresponding to 411 different animals, were obtained. Of
them, 188 were captured at least twice, and at most 10 times. The probability
of recapture depends on sex and age group, as shown. It suggests an increase of
the probability of recapture with age, independently supported by the
probability of recapture as a function of weight (not shown). J: juveniles, SA:
sub-adults, A: adults, F: females, M: males.} \label{captured}
\begin{tabular*}
{\hsize}{l@{\extracolsep{0ptplus1fil}}r@{\extracolsep{0ptplus1fil}}r@
{\extracolsep{0ptplus1fil}}r@{\extracolsep{0ptplus1fil}}r}\hline \hline
        & J  & SA & A   & Total \\ \hline
Captured only once: & & & & \\
  F     & 7  & 13 & 98  & 118 \\
  M     & 8  & 18 & 79  & 105 \\
  Total & 15 & 31 & 177 & 223 \\ \hline
Recaptured at least once: & & & & \\
  F     & 2  &  2 &  71 & 75  \\
  M     & 0  & 16 & 97  & 113 \\
  Total & 2  & 18 & 168 & 188 \\ \hline
Probability of recapture: \\
  F     & 0.22 & 0.13 & 0.42 \\
  M     & 0.00 & 0.47 & 0.55 \\
  Total & 0.13 & 0.37 & 0.49 \\
\hline \hline
\end{tabular*}
\end{table}

The AK model~\citep{ak,akpy} is based on a fundamental set of biological
features that characterize the transmission of Hantavirus among rodent
populations, and involves five parameters: the birth rate $b$ of the rodents
(mice), their death rate $c$, the environmental parameter $K$, the contagion
rate $a$, and the diffusion coefficient $D$. In terms of these, the mice
populations $M_s$ (susceptible) and $M_i$ (infected) obey
\begin{eqnarray}
\frac{\partial M_s}{\partial t} &=&
b (M_s+M_i)-cM_s-\frac{M_s (M_s+M_i)}{K(x,t)}-aM_s M_i+D\nabla^{2} M_s, \label{ak1}\\
\frac{\partial M_i}{\partial t} &=& -cM_i-\frac{M_i (M_s+M_i)}{K(x,t)}+aM_s
M_i+D\nabla^{2} M_i.
\label{ak2}
\end{eqnarray}
All the parameters except $K$ are considered to be independent of time $t$ and
space $x$.

The measurement of the birth and death rates $b$, $c$ presents no  special
challenge. The environmental parameter $K$ is sometimes measured from food and
vegetation measurements, and also often obtained from aerial photographs of the
landscape. Its variation in time and space can be  well characterized although
its absolute values are difficult to obtain. The encounter infection rate $a$
is notoriously hard to measure from  observations of individual mouse-mouse
interactions \citep{botten2002}. At this stage  of observational technique, we
must assume that it is a floating parameter.

The last of the five parameters, the diffusion constant $D$, is crucial to the
AK description since the assumed mechanism for the spread of the  epidemic is
the diffusion (movement) of infected mice over the terrain followed by the
transmission of infection to susceptible mice. In principle, it appears
straightforward to measure $D$ from records of mice movement in a
mark-recapture experiment, as was done by \citet{ovaskainen04} in a recent
study on the dispersal of butterflies in a heterogeneous habitat. The
investigation reported in the present paper began as an attempt to extract $D$
directly in that manner. We will see that examination of the data has indeed
allowed us to obtain values of $D$ but also led us to a number of important
conclusions about the transmission of Hantavirus among rodent populations, and
has suggested substantial changes to be introduced into the theory of the
spread of Hantavirus.

\section{The data set: \emph{Zygodontomys brevicauda} in Panama}

At the beginning of 2000, human cases (more than 20 in the first cluster) of
Hantavirus Pulmonary Syndrome (HPS) were recognized from the Azuero Peninsula,
Panama. Following the outbreak, it was discovered that the pigmy rice rat
\emph{Oligoryzomys fulvescens} and the short-tailed cane mouse
\emph{Zygodontomys brevicauda} harbored two novel hantaviruses, Choclo virus
(responsible for the HPS cases), and Calabazo virus (not known to cause human
disease) respectively~\citep{vincent2000}. The data set we have selected in
this study was obtained as the result  of a mark-recapture observation
performed in the Azuero Peninsula, in Tonos\'{\i} (Los Santos), Panama, from June 27
to November 20, 2003. The observation corresponds to the rainy season in the
region, and following several years of relative draught.
During these years, clinical cases of HPS were rare (after being more frequent
in 1999-2001) and rodent densities were probably relatively low.

Measurements were made on several species of rodents, of which we choose for
the purposes of the present paper \emph{Z. brevicauda}, host of Hantavirus
Calabazo. This choice was made because \emph{Z. brevicauda} was the most
abundant species in the field study. A summary of the characteristics of the
populations is presented in Table~\ref{captured}. The data relevant to the
analysis of the movement consist of the position and time of capture of those
mice that are captured at least twice, thus allowing for the calculation of
their displacements from one location to another. The measurements were made
with a square array of $7\times 7$ Sherman traps, separated 10 m  each. Each
measuring session lasted 3 days, with a time of recurrence to the  same site of
about one month. The number of trapping grids used in the  study was 24, in 4
different sites. Each grid was set up across the edge between forest and
pastures~\citep{suzan04}. For our analysis given below, all grids have been
rotated such that the edge runs along the $y$ direction.

\begin{figure}[t]
\centering  \resizebox{10cm}{!}{\includegraphics{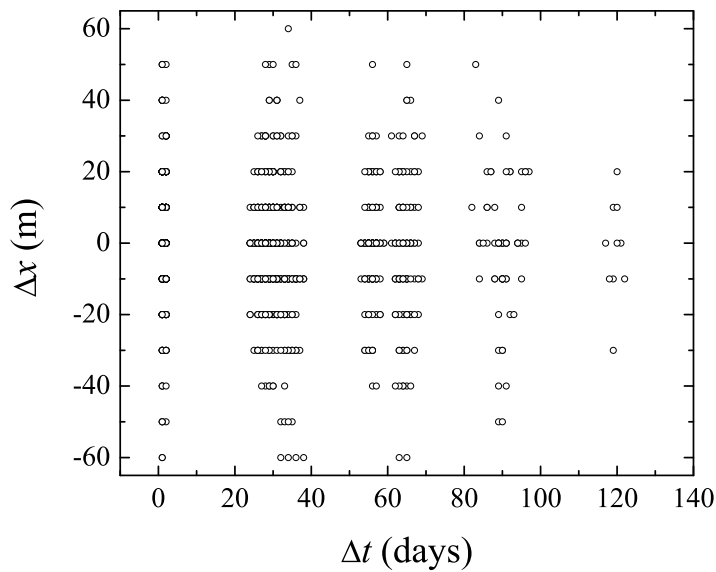}}
\caption{Displacements of the recaptured mice, projected along the $x$
direction which is the direction into and out of the forest relative to the
pastures in the landscape. Each point corresponds to a displacement of one of
the 411 mice that were captured  more than once. Some mice were captured more
than twice, thus contributing with more than one point to  this set.}
\label{displacements}
\end{figure}

\section{The movement of \emph{Z. brevicauda}}

A few animals were captured a sufficient number of times (about 10), during a
period of months, as to allow us to form a useful picture of the mouse walks.
Unfortunately, there are only five of these mice. The limited number makes it
impossible to carry out any statistical analysis of the properties of the
walks.

Despite this drawback, we were able to proceed with the analysis  because the
data set contains hundreds of recapture events, corresponding to
\emph{different} mice, each one providing us with a  displacement at a certain
time scale. These time scales are 1 and 2 days (if the  recapture occurs during
the same session, that lasts three days); about 1 month (if the recapture
occurs at the next session); and about 2 months (at the second next session), 3
months, and 4 months, respectively. At progressively  longer time scales there
are, certainly, less recapture events; but there are a sufficient  number of
them for sensible results to be obtained up to the 3-month scale. The data are
shown in Fig.~\ref{displacements}, for the $x$-component of the 2-dimensional
displacements of the mice. Each point in the graph represents a displacement
$\Delta x$ taking place during an interval $\Delta t$. Each displacement
corresponds to the movement of a single mouse, but different displacements may
or may not correspond to the same animal. We consider these  displacements as
our elementary events. Furthermore, we specifically assume them to  be
statistically independent. There are repeated events in the set, namely
displacements with the same $\Delta x$ and $\Delta t$, a fact  not represented
in Fig.~\ref{displacements}. In the plot, it can be observed that, besides the
1- and 2-day intervals, the data are scattered in ``clouds'' around 1, 2, 3 and
4 months. The reason for this is twofold. First, each session consists of 3
consecutive days of capture; therefore, the interval between two recaptures in
different sessions is not a constant number of days.  Second, variations due to
the logistics of field work result in the time between sessions not being
precisely 30 days.

On each time scale, the set of available displacements is taken to represent a
statistical ensemble, i.e., a population of ideal mice with  certain
statistical properties. Also, the displacements measured on each time scale
correspond to a progressively coarser graining of the actual mice walks,
containing an indeterminate (but presumably large) number of steps already on
the 1-day scale.

\begin{figure}[t]
\centering \resizebox{10cm}{!}{\includegraphics{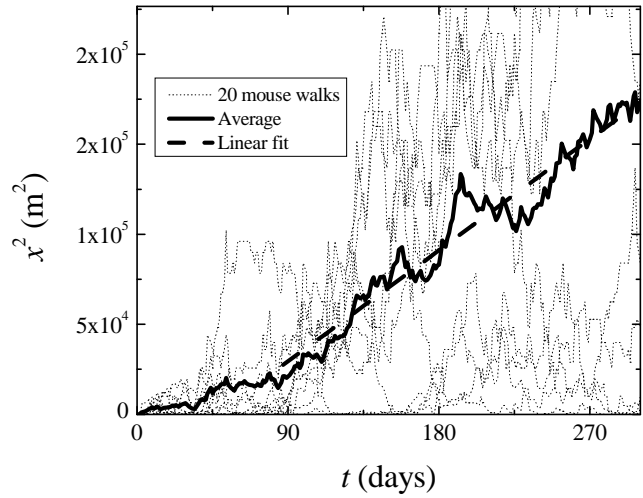}}
\caption{Square of the displacement as a function of time, on the time scale of
1 day, for 20 artificial ``mouse walks'' based on the recorded displacements.
The heavy line is an ensemble average.} \label{d1day}
\end{figure}

Using available data on each time scale, we construct mouse walks by randomly
shuffling the displacements. These walks represent  instances of possible
walks, on that time scale, with exactly the same statistical properties as a
hypothetical ``representative mouse". We construct  various such walks and
perform ensemble averages to obtain the mean square displacement as a function
of time. For example, on the time scale of 1 day, 20 instances  of the walk
produce the result shown in Fig.~\ref{d1day}. Diffusive behavior is inferred
from the clearly linear rise of the mean square displacement. On longer time
scales the same analysis produces smaller diffusion coefficients. However,
application of the  method is not reliable on these longer scales because it
does not  permit a correct evaluation of errors.

\begin{figure}[t]
\centering \resizebox{10cm}{!}{\includegraphics{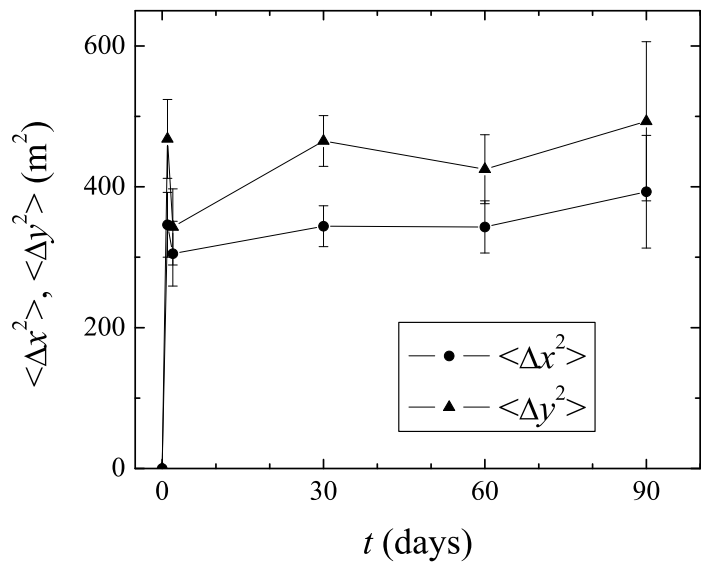}}
\caption{Mean square displacement as a function of time, using the different
time scales available from the data of Fig.~\ref{displacements}. The two curves
correspond to the two directions in space.}  \label{xsq}
\end{figure}

The above considerations clarify the underlying diffusive nature of mice
motion. A direct analysis of the probability distribution function $P(\Delta
x)$ is possible, also on each time scale. $P(\Delta x)$ is a bell-shaped
distribution, with a well-defined variance that can be used  to characterize
the evolution of the mean square displacement of the representative mouse as a
function of time. The result of this analysis is  in Fig.~\ref{xsq}, where we
show the mean square displacements in both the $x$ and the $y$ directions.
Unlike in a simple diffusive process, the observed mean square displacement is
observed to saturate to a value near $400$ m$^2$. The saturation value  is
different for  $\langle\Delta x^{2}\rangle$ and for $\langle\Delta y^{2}\rangle$,
indicating an anisotropy in the system. We have already  mentioned above that
each trapping grid contains an edge between forest and pastures, mainly along
the $y$  direction. This is surely responsible for the anisotropy.

\section{Movement in Confined Spaces: Rodent Home Ranges and Grid Sizes}

The saturation of the mean square displacement, clear from the observations
discussed above, implies the  existence of a spatial scale in the rodent
system. One obvious possibility is  that  the saturation is a manifestation of
the confined motion of the rodents, i.e., the   existence of home ranges
\citep{Burt}. In order to treat the problem quantitatively, we first present a
simple  calculation of the effect of confinement on the mean square
displacement of a random walker.

\subsection{Effect of Confinement on $\langle x^{2}\rangle$ }

There are several ways one might model the effect of confinement on the motion
of rodents. One is to treat the motion as obeying, not a  diffusion equation,
but a Fokker-Planck equation in an attractive  potential \citep{kk}. A simpler
way, which we adopt here, is to consider  the motion as occurring via simple
diffusion but confined to a box  whose size represents the home range.
Restricting the analysis to  1-dimension for simplicity, we solve the diffusion
equation for the probability per unit length $P(x,t)$ of finding the mouse at
position  $x$ and time $t$ inside a bounded domain of length $L$. Taking into
account
the symmetry of the problem, the general solution can be written as $%
P(x,t)=A_{0}+\sum{A_{\lambda }\cos \left[ \lambda x\right] e^{-\lambda
^{2}Dt}}$. By imposing the condition that a mouse cannot escape from the home
range (zero flux boundary condition), the allowed $\lambda $ coefficients can
be calculated and  the solution written as
\begin{equation}
P(x,t)=\frac{1}{L}+\frac{2}{L}\sum_{n=1}^{+\infty }A_{n}\cos \left[ \frac{%
2n\pi x}{L}\right] e^{-\frac{\left( 2n\right) ^{2}\pi ^{2}Dt}{L^{2}}},
\label{gensol}
\end{equation}
wherein
\begin{equation}
A_{n}=\frac{2}{L}\int_{-L/2}^{L/2}dx\cos \left[ \frac{2n\pi x}{L}\right]
P(x,0).  \label{Acoeff}
\end{equation}
With the  initial condition that the mouse is within a length $\alpha$ centered
around the origin of the box, specifically
\begin{equation}
P(x,0)=\left\{
\begin{array}{ll}
1/\alpha & \left| x\right| \leq \alpha/2 \\
0        & \left| x\right| > \alpha/2
\end{array}
,\right.  \label{initcond}
\end{equation}
the evolution at all times can be written as
\begin{equation}
P(x,t)=\frac{1}{L}+\frac{2}{\alpha \pi }\sum_{n=1}^{+\infty }\frac{\sin
\left( \frac{n\pi \alpha }{L}\right) \cos \left( \frac{2n\pi  x}{L}\right) }{n%
}e^{-\frac{\left( 2n\right) ^{2}\pi ^{2}Dt}{L^{2}}}.  \label{confinedP}
\end{equation}
The initial condition (\ref{initcond}) reduces to a $P(x,0)=\delta (x)$ in the
limit $\alpha \rightarrow 0$. The calculation of the mean  square displacement
$\langle x^{2}\rangle =\int_{-L/2}^{L/2}dxx^{2}P(x,t)$ gives
\begin{equation}
\langle x^{2}\rangle =\frac{L^{2}}{12}\left\{  1-\frac{L}{\alpha \pi
^{3}}\sum_{n=1}^{+\infty }\frac{\left( -1\right) ^{n+1}\sin \left(  \frac{ n\pi
\alpha }{L}\right) }{n^{3}}e^{-\frac{\left( 2n\right) ^{2}\pi  ^{2}Dt}{
L^{2}}}\right\} .
\label{confineddisp}
\end{equation}
We see that $\langle x^{2}\rangle \simeq \alpha ^{2}/12+2Dt$ for $t\rightarrow
0$. We also see that $\langle  x^{2}\rangle$ saturates to $L^{2}/12$ for
$t\rightarrow +\infty$. If $P(x,0)=\delta (x)$, the mouse will move initially
as if no home range existed. However, its diffusive motion will be limited in
extent by the presence of the home range and $\langle x^{2}\rangle $ will
eventually saturate. The 1-dimensional calculation captures the essential
features. The 2-dimensional extension, appropriate to mouse  movement on the
terrain, is straightforward to obtain because $P(x,y,t)$
is given by the product of two functions of the form~(\ref{confinedP}), one for the $x$%
-plane with $D_{x}$ and $L_{x}$ and the other for the $y$-plane with $D_{y}$
and $L_{y}$. Allowing for the fact that differences in terrain could be
reflected in the differences in $D_{x}$ and $D_{y}$ (given that the  traps are
laid out systematically relative to `edges' in the landscape  between forests
and pastures), we give a usable expression for   the  average mean square
displacement in 2-dimensions which can be used  directly for the interpretation
of the observations.
\begin{eqnarray}
\langle x^{2}\rangle &+&\langle y^{2}\rangle  = \frac{%
L_{x}^{2}+L_{y}^{2}}{12} \left[1-\frac{1}{\alpha \pi ^{3}}\sum_{n=1}^{+\infty }\frac{%
\left( -1\right) ^{n+1}}{n^{3}}\times \right. \nonumber \\
&\times& \left. \left\{ L_{x}\sin \left( \frac{n\pi  \alpha }{%
L_{x}}\right) e^{-\frac{\left( 2n\right) ^{2}\pi ^{2}D_{x}t}{L_{x}^{2}}%
}+L_{y}\sin \left( \frac{n\pi \alpha }{L_{y}}\right) e^{-\frac{\left( 2n\right)
^{2}\pi ^{2}D_{y}t}{L_{y}^{2}}}\right\}\right] .
\end{eqnarray}

Application of this analysis to the mouse data is straightforward. The
short-time part of the mean square displacement gives the diffusion constant,
averaged over the two directions, to be $200\pm 50$ m$^2$/d. The saturation
value appears to imply that the  home range $L$ equals about 70 m. While these
initial considerations  suggest that the Panama data allow us to confirm the
existence of  rodent home ranges, as well as to measure their extent, careful
observation introduces a note of caution: we notice that the derived value of
the home range is of the order of the size of the measurement grid $G=60$ m.
Could sampling from a limited domain in space lead to saturation and mislead
one into drawing  incorrect conclusions about the home range? To answer this
question we  carry out the following analysis.

\subsection{Effect of Limited Spatial Observations on $\langle x^{2}\rangle$}

Consider the motion of a random walker in unbounded space (no home ranges) but
let the mean square displacement be calculated from observations in a
\emph{limited} part of space of  size $G$ placed symmetrically around the
origin for simplicity. We  obviously  have
\begin{equation}
\langle x^{2}\rangle  =\frac{\int_{-G/2}^{G/2}dx\;x^{2}P\left( x,t\right)
}{\int_{-G/2}^{G/2}dx\;P\left( x,t\right) }
\end{equation}
where the probability $P(x,t)$ is given by the propagator of the diffusion
equation in \emph{unbounded} space:
\begin{equation}
P\left( x,t\right) =\frac{e^{-\frac{x^{2}}{4Dt}}}{\sqrt{4\pi Dt}}.
\end{equation}
Substitution gives
\begin{equation}
\langle x^{2}\rangle =2Dt\left[ 1-\frac{Ge^{-\frac{G^{2}}{16Dt}}}{\sqrt{4\pi
Dt}\erf\left(
\frac{G}{4\sqrt{Dt}}\right) }\right] .
\label{erff}
\end{equation}
At short times $\left\langle x^{2}\right\rangle \simeq 2Dt$ while at  long
times it saturates to the value $G^{2}/12$. The 2-dimensional result is
trivially obtained as a generalization of (\ref{erff}).

Note that the behavior of a system without home ranges but with a finite window
of observation is qualitatively different from that of a system with home
ranges and an infinitely large window. The mean square displacement in the
former (Eq.~(\ref{erff})) has quite a different time dependence from the latter
(Eq.~(\ref{confineddisp})), since the error function expressions differ
considerably from exponentials. It is easy to see that in both  cases the mean
square displacement starts out at short times as $2Dt$ and saturates to a
constant at long times. The saturation is to  $G^2/12$ in the first case and to
$L^2/12$ in the second. There is thus  potential for confusion. One could
mistakenly interpret what is  actually the measurement grid size $G$ to be the
rodent home range $L$.
\begin{figure}
\centering
\resizebox{10cm}{!}{\includegraphics{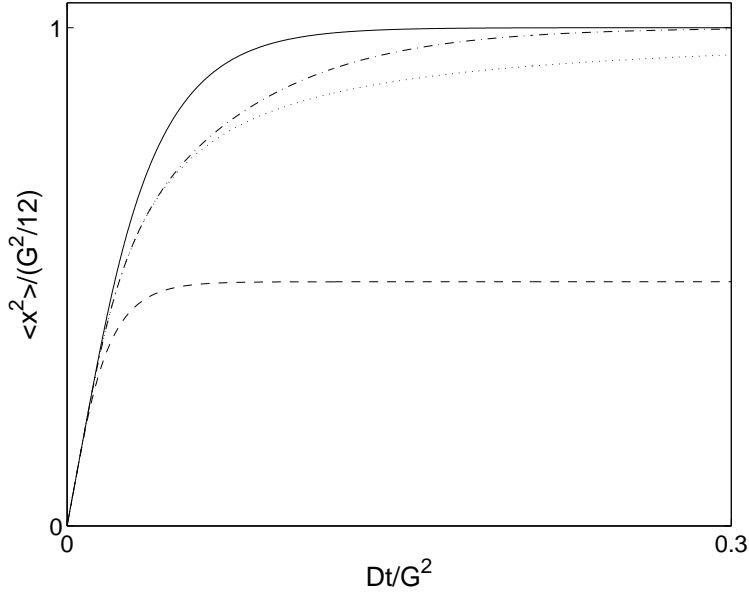}}
\caption{Mean square displacement in units of $G^{2}/12$ as a function of time in
units of $G^{2}/D$ for different values of the ratio $G/L$, i.e., the ratio of the
measuring grid divided by the width of the home range. The dashed, solid,
dash-dotted and dotted lines correspond, respectively, to $G/L=$1.43, 1, 0.7,
0. When $G>L$, $\langle x^{2}\rangle$ is seen to saturate to the lower value
$L^{2}/12$, whereas for all other ($G\le L$) cases, it saturates to $G^{2}/12$. The
time dependence differs qualitatively for the various cases. The difference
might be discernible---in principle---with more refined measurements.}
\label{msdcomp}
\end{figure}

Combination of the two elements discussed above, the home range effect and the
grid size effect, is straightforward to analyze. Consider the  situation in
which a mouse is moving randomly inside a home range of width $L$ but  is
observed only inside a region of width  $G$, both concentric for simplicity. If
the grid is larger than the  home range ($G>L$), the evolution of the mean
square displacement is exactly given by the  previous result
(\ref{confineddisp}). In the other case, ($G<L$), the evolution is
\begin{multline}
\langle x^{2} \rangle =
\frac{G^{2}}{12} \times \\
\left\{
       \frac
       {
        \zeta + \frac {6L} {\alpha\zeta^{2}\pi^{4}}
        {\displaystyle \sum_{n=1}^{+\infty }}
        \sin \left( \frac {n\pi\alpha} {L} \right)
        \left\{
               2\pi\zeta \frac {\cos \left( n\pi\zeta \right)} {n^{3}}
               + \sin \left( n\pi\zeta \right)
             \left[
                   \frac {\pi^{2}} {n^{2}} \zeta^{2}- \frac {2} {n^{4}}
             \right]
        \right\}
        e^{-\frac {\left( 2n \right)2\pi^{2}Dt} {L^{2}}}
       }
       {
        \zeta+\frac {2L} {\alpha\pi^{2}}
        {\displaystyle \sum_{n=1}^{+\infty }}
        \frac {
               \sin \left( \frac {n\pi\alpha } {L}
                    \right)
               \sin \left( n\pi\zeta
                    \right)
              }
              {n^{2}}
        e^{
           -\frac {\left(2n\right)^{2}\pi^{2}Dt} {L^{2}}
          }
       }
\right\},
\label{obsmsqdhr}
\end{multline}
wherein $\zeta=G/L$. Equation (\ref{obsmsqdhr}) reduces to Eq.
(\ref{confineddisp}) when $\zeta=1$. It can be shown that $\langle
x^{2}\rangle$ goes as $\alpha ^{2}/12+2Dt$ for $t\rightarrow 0$ and saturates
to $L^{2}/12$ for $t\rightarrow +\infty $. In Fig.~(\ref{msdcomp}) we compare
Eqs.~(\ref{confineddisp}),~(\ref{erff}) and~(\ref{obsmsqdhr}) for an initially
localized condition $P(x,0)=\delta (x)$. Despite the fact that all of the three
curves (with $G<L$) show the same linear behavior at short times and reach the
same saturation value, they differ considerably at intermediate times. A fourth
curve (dashed, lowermost in the group) for which $G>L$ is also shown in the
same graph.
\begin{figure}
\centering
\resizebox{10cm}{!}{\includegraphics{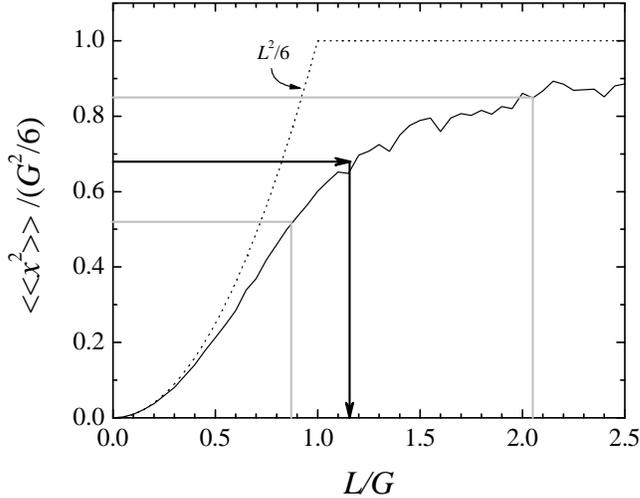}}
\caption{Mean square displacement in units of $G^{2}/6$ as a function of the ratio
of the home range size to the grid size, $L/G$. The curve is the result of a
simulation of $105$ steps for each value of $L/G$. The arrow shows the average
(in both directions $x$ and $y$) mean square displacement observed in the
measurement, from the data shown in Fig.~\ref{xsq}, and the inferred value of
the home range ($L/G=1.16$). The grey lines show the error bounds. The dotted
lines show the analytic result derived in the Appendix, valid for home ranges
concentric with the grid.}
\label{simul}
\end{figure}

An important step in the interpretation of saturation data is the verification
of the sensitivity of the theoretical prediction to the initial position of the
mice, as well as to the position of the home ranges with respect to the grid.
In the Appendix we provide further details on the calculation of the mean
square displacement, relevant to the dependence on the distribution of initial
locations of the mice. To assess the combined effect, we have done a numerical
experiment in which we measure simulated mice displacements. Each mouse is
supposed to occupy uniformly its own home range, and all the home ranges are
uniformly distributed in space. We limit the calculation of the mean square
displacement to a window of size $G$, thus simulating the conditions of the
measurement in the field. The result is shown in Fig.~\ref{simul}, where the
normalized mean square displacement is plotted as a function of the normalized
home range size $L/G$. The available data allow us to conclude that the home
range of \emph{Z. brevicauda} in Panama (averaged in the two directions) is
$70$ m, with asymmetric error bounds from $50$ to 120 m.

\section{Discussion}

Extraction of quantitative information concerning the parameters in the
A\-bram\-son-Kenkre theory \citep{ak} of Hantavirus spread was the initial task
undertaken in the present investigation. This was prompted by the success of
that theory in reproducing qualitatively observed features  such as the
sporadic disappearance of infection during time periods of  low carrying
capacity, and the existence of refugia where infection  persists and from where
it propagates in the form of waves when  conditions become favorable for such
propagation. The key to the AK  description of these processes was a
transcritical bifurcation  predicted by the theory and controlled by the
carrying capacity.  Crucial to quantitative application of the theory is the
diffusion  constant of the rodents, and the first aim of the present
investigation  was to extract this diffusion constant from movement
observations  collected for \emph{Zygodontomys brevicauda} in the peninsula de
Azuero  in Panama in a 5-month period in 2003. While logistic growth which is
also part of the theory of ref \citep{ak} is generally considered well
established, both from field and laboratory studies \citep{murray}, equally
strong justification for assuming diffusive transport for rodent  movement is
not available. The goal of extracting the diffusion  constant has been met. The
result particular to the species and  location considered is $D=200 \pm 50$
m$^2$ per day.

During the process of this extraction, we encountered a spatial scale  in the
rodent system which could be representative of the home range of  the mice.
Preliminary investigations showed that the observed spatial scale could be
reflecting the size of the measuring grid instead of a characteristic of the
rodent system. Since accuracy of interpretation will be considerably increased
by having multiple values of the grid size, our analysis underscores the
importance for additional observations to be undertaken with grids of varying
size. Motivated by this idea, we carried out further developments of the our
theory which resulted in the simulation curve shown in Fig.~\ref{simul}. The
mean value of the home range for the two directions can be read off from the
plot to be $L_{x}$=60 m and $L_{y}$=90 m.

Because our prescription for extracting the diffusion constant relies  on the
initial time evolution of the mean square displacement rather  than its
saturation value, we have been successful in obtaining a  usable estimate of
the diffusion constant. However, even that  evaluation suffers from the fact
that reconstruction of mouse walks  from observations is difficult because
multiple measurements for  individual mice are extremely rare. Therefore, we
suggest strongly that  systematic measurements of rodent movement be undertaken
in other ways  such as radio telemetry to confirm and sharpen the estimates of
the diffusion constant we have obtained in the present analysis. Elsewhere we
will present our separate analysis \citep{adgkdppy} of mice movement  data
taken from web measurements~\citep{parmenter03}.

The present analysis has pointed to the importance of the concept of  home
ranges, which have been known and discussed in many places in the literature
earlier \citep{Burt,anderson82,ford79}. They would generally be reflected in
animal movement measurements in the way we have detailed  in our calculations
in the present paper and could additionally be of  crucial importance in the
theory of the spread of epidemics. While they  do not appear in the AK
description (the tacit assumption in that  formalism being that they are larger
than other lengths of interest),  we have now developed \citep{kgac} a series
of models to treat them  explicitly. The basic idea in our new theory is to
consider the  dynamics of two types of mice, stationary and itinerant (and
susceptible and infected in each category). The stationary mice are the adults
that move within their home ranges and do not stray far from the burrow. The
itinerant mice are the subadults that must leave to find  their own home
ranges. Our studies employ a combination of nonlinear  analysis and
simulations. The results will be presented elsewhere.

\begin{ack}
It is a pleasure to thank Robert Parmenter for numerous discussions. We also
thank Instituto Conmemorativo Gorgas, Drs. Blas Armien and An\'{\i}bal Armien, for
generously making this work possible. We are grateful to Armando McKency, Mario
\'{A}vila, Omar Vargas, Nelson R\'{\i}os, Francisco Crespo, Ernest Valdez, Marjorie
Hudson, Ricardo Jim\'{e}nez and Eustiquio Broce  for splendid field assistance and
other Instituto Gorgas staff for general support. The field work of Gerardo
Suz\'{a}n, Erika Marc\'{e}, and Terry Yates was funded by NSF Dissertation Improvement
Grant 3-25471-3100. Guillermo Abramson acknowledges partial funding by CONICET
(PEI 6482), and by Fundaci\'{o}n Antorchas. This work was supported in part by
DARPA under grant no. DARPA-N00014-03-1-0900, by NSF/NIH Ecology of Infectious
Diseases under grant no. EF-0326757, and by the NSF under grant no.
INT-0336343.
\end{ack}

\appendix

\section{Dependence of $\langle x^{2}\rangle$ on  distribution of
initial location of rodents within the grid}

The mean square displacement obviously depends on the initial location of the
mice, $x_{0}$. Let us consider the case of infinite home range,  but finite
$G$, and determine the effect on (\ref{erff}) of the  distribution
$\mathcal{P}(x_{0})$ of initial locations of the mice. We thus evaluate
$\langle (x-x_{0})^{2}\rangle$ and calculate its average, denoted by an additional
$\langle\cdots\rangle$, with respect to the distribution $\mathcal{P}(x_{0})$
of the initial positions through $\langle \langle (x-x_{0})^{2}\rangle\rangle
=\int_{-G/2}^{G/2}dx_{0}\mathcal{P}(x_{0})\langle (x-x_{0})^{2}\rangle $.
Proceeding as in Eq.~(\ref{erff}), we  have
\begin{multline}
\langle\langle (x-x_{0})^{2}\rangle\rangle  = \\
2Dt-2Dt\int_{-G/2}^{G/2}dx_{0}{\mathcal{P}}(x_{0})\left\{  \frac{ \left(
\frac{G}{2}+x_{0}\right) e^{-\frac{\left(  \frac{G}{2}+x_{0}\right)
^{2}}{4Dt}}+\left( \frac{G}{2}-x_{0}\right) e^{-\frac{\left(  \frac{G}{2}
-x_{0}\right) ^{2}}{4Dt}}}{\sqrt{\pi Dt}\left[ \,\erf\left(  \frac{
\frac{G}{2}+x_{0}}{\sqrt{4Dt}}\right) +\,\erf\left(  \frac{\frac{G}{2}
-x_{0}}{\sqrt{4Dt}}\right) \right] }\right\} +  \\
+4\sqrt{\frac{Dt}{\pi }}\int_{-G/2}^{G/2}dx_{0}{\mathcal{P}}(x_{0})x_{0}
\left\{
       \frac{e^{-\frac{\left( \frac{G}{2}+x_{0}\right) ^{2}}{4Dt}}-e^{-\frac{\left(  \frac{
       G}{2}-x_{0}\right) ^{2}}{4Dt}}}{\,\erf\left( \frac{\frac{G}{2}+x_{0}}{
       \sqrt{4Dt}}\right) +\,\erf\left( \frac{\frac{G}{2}-x_{0}}{\sqrt{4Dt}}
       \right) }
\right\}.
\label{avepopmsdnohr}
\end{multline}
This reduces to Eq.~(\ref{erff}) when $\mathcal{P}(x_{0})=\delta (x_{0})$. The
evolution at short times is found to be independent of $\mathcal{P}(x_{0})$:
\begin{equation}
\lim_{t\rightarrow 0}\langle\langle (x-x_{0})^{2}\rangle\rangle \simeq 2Dt,
\label{shorttimenohr}
\end{equation}
while the saturation turns out to depend on $\mathcal{P}(x_{0})$ and is given
by
\begin{equation}
\lim_{t\rightarrow +\infty }\langle\langle (x-x_{0})^{2}\rangle\rangle =
G^{2}/12+\int_{-G/2}^{G/2}dx_{0}\mathcal{P}(x_{0})x_{0}^{2}.
\label{longtimenohr}
\end{equation}
Notice that for a uniform initial distribution $\mathcal{P}(x_{0})=1/G$, the
integral equals $G^{2}/12$, and the saturation value  is $G^{2}/6$. This value
is twice that obtained  for the case of initial placement of the mice at the
center of the  grid. We see from  (\ref{longtimenohr}) that its right hand side
lies  always between $G^{2}/12$ and $G^{2}/3$ depending on the  initial
distribution.

The time-dependent evolution of $\langle\langle (x-x_0)^{2}\rangle\rangle$ can
also be determined when the home range is not infinite. If $x_0$ is not the
center of the home range, $P(x,t)$ is no longer symmetric with respect to the
origin, and the series solution for $P(x,t)$ now contains sine (in addition to
cosine) functions:
\begin{multline}
P_{x_{0}}(x,t) = \\
\frac{1}{L}+\frac{2}{\alpha \pi }\sum_{n=1}^{+\infty  }
\frac{\cos \left( \frac{n\pi 2x_{0}}{L}\right) \sin \left( \frac{n\pi \alpha
}{L}\right) \cos \left( \frac{2n\pi x}{L}\right) }{n}e^{-\frac{\left(
2n\right) ^{2}\pi ^{2}Dt}{L^{2}}} + \\
+\frac{4}{\alpha \pi }\sum_{n=1}^{+\infty }\frac{\sin \left(  \frac{\left(
2n-1\right) \pi x_{0}}{L}\right) \sin \left( \frac{\left( 2n-1\right)  \pi
\alpha }{2L}\right) \sin \left( \frac{\left( 2n-1\right) \pi  x}{L}\right) }{
2n-1}e^{-\frac{\left( 2n-1\right) ^{2}\pi ^{2}Dt}{L^{2}}}.
\end{multline}
Here the initial probability is
\begin{equation}
P_{x_{0}}(x,0)=\left\{
\begin{array}{ll}
1/\alpha & \left| x-x_{0}\right| \leq \alpha/2 \\
0        & \left| x-x_{0}\right| > \alpha/2
\end{array}
.\right.
\end{equation}
We continue to consider the home range and grid window to be concentric for
simplicity. The calculation follows the steps shown in earlier  cases and
yields, for $\langle\langle (x-x_{0})^{2}\rangle\rangle$, the expression:
\begin{multline}
\int_{-G/2}^{G/2}dx_{0}{\mathcal P}(x_{0})x_{0}^{2}+\frac{G^{2}}{12}\int_{-G/2}^{G/2}dx_{0}{\mathcal P}(x_{0}) \times \\
\shoveleft{
\times\left\{ \zeta  +\frac{6L}{\zeta ^{2}\alpha \pi
^{4}}\sum_{n=1}^{+\infty }\cos \left( \frac{n\pi 2 x_{0}}{L} \right) \sin
\left( \frac{n\pi \alpha }{L}\right) \right. \times
}\\
\shoveright{
\times \left[
      2\pi\zeta \frac{
                      \cos \left( n\pi \zeta \right)
                     }
                     {n^{3}}
      +\sin \left( n\pi \zeta \right)
      \left(
            \frac{\pi^{2}} {n^{2}} \zeta^{2} - \frac{2} {n^{4}}
      \right)
\right] e^{-\frac{\left( 2n\right) ^{2}\pi  ^{2}Dt}{L^{2}}} +
}\\
\shoveleft{
+\frac{96}{\zeta ^{2}\pi ^{2}}\frac{x_{0}}{\alpha
}\sum_{n=1}^{+\infty }\frac{ \sin \left( \frac{\left( 2n-1\right) \pi
x_{0}}{L}\right) \sin \left(
\frac{
\left( 2n-1\right) \pi \alpha }{2L}\right) }{\left( 2n-1\right) ^{2}} \times
}\\
\shoveright{ \left.\times \left[
      \zeta \cos \left( \frac{\left(2n-1\right)\pi\zeta} {2}
                 \right)
      -\frac{2}{\pi}
      \frac{\sin \left(
                       \frac{\left( 2n-1\right)\pi\zeta}{2}
                 \right)
           }
           {\left( 2n-1\right)}
\right] e^{-\frac{\left( 2n-1\right)  ^{2} \pi ^{2}Dt}{L^{2}}}\right\}
}\\
\times\left\{ \zeta +\frac{2L}{\alpha \pi ^{2}} \sum_{n=1}^{+\infty
}\frac{\cos \left( \frac{n\pi 2x_{0}}{L}\right) \sin \left(
\frac{n\pi \alpha }{L}\right) \sin \left( n\pi \zeta \right)  }{n^{2}}
e^{-\frac{\left( 2n\right) ^{2}\pi ^{2}Dt}{L^{2}}}\right\} ^{-1}.
\label{avepopmsdhr}
\end{multline}

The time dependence of Eq.~(\ref{avepopmsdhr}) at short and long times can be
shown to tend to that given by Eq.~(\ref{shorttimenohr}) and
(\ref{longtimenohr}), respectively, for the case of free diffusion. In the case
$L<G$, the result is simply Eq.~(\ref{avepopmsdhr}) with the parameter $\zeta$
set equal to 1 and $G$ set equal to $L$. In such a case the saturation value is
given by an expression similar to Eq.~(\ref{longtimenohr}) but with $L$
replacing $G$ everywhere.

\end{document}